\documentclass[11pt,reqno]{article}

\RequirePackage{amsthm,amsmath,amsfonts,amssymb}
\RequirePackage[authoryear]{natbib}
\RequirePackage[colorlinks,citecolor=blue,urlcolor=blue]{hyperref}
\RequirePackage{graphicx}
\usepackage{caption}
\usepackage{subcaption}
\usepackage{amsmath, amssymb, amscd, amsthm, amsfonts}
\usepackage{graphicx}
\usepackage[perpage,symbol*]{footmisc}
\usepackage{float}
\usepackage{hyperref}
\usepackage{pgfplots}
\usepackage{mathtools}
\usepackage{color}  
\usepackage{tikz}  
\usepackage{quiver}
\usepackage{bbm}

\newcommand{\commentout}[1]{}

\usetikzlibrary{shapes, arrows, calc, arrows.meta, fit, positioning} 
\tikzset{  
    -Latex,auto,node distance =1.5 cm and 1.3 cm, thick,
    state/.style ={ellipse, draw, minimum width = 0.9 cm}, 
    point/.style = {circle, draw, inner sep=0.18cm, fill, node contents={}},  
    bidirected/.style={Latex-Latex,dashed}, 
    el/.style = {inner sep=2.5pt, align=right, sloped}  
}  

\usepackage{authblk}

\usepackage{algorithm,algcompatible,amsmath}
\algnewcommand\INPUT{\item[\textbf{Input:}]}%
\algnewcommand\OUTPUT{\item[\textbf{Output:}]}%

\usepackage{enumitem}

\usepackage[english]{babel}
\setcitestyle{authoryear,open={(},close={)}}

\usepackage{colortbl,dcolumn}

\usepackage{listings} 
\usepackage{xcolor} 

\renewcommand{\raggedright}{\leftskip=0pt \rightskip=0pt plus 0cm}
\raggedright

\def\hilite<#1>{%
	\temporal<#1>{\color{blue!25}}{\color{magenta}}%
	{\color{blue!55}}}

\newcolumntype{H}{>{\columncolor{blue!20}}c!{\vrule}}
\newcolumntype{H}{>{\columncolor{blue!20}}c}

\numberwithin{equation}{section} 
\numberwithin{theorem}{section}
\numberwithin{lemma}{section} 
\numberwithin{corollary}{section}
\numberwithin{definition}{section}
\numberwithin{proposition}{section} 
\numberwithin{remark}{section}
\numberwithin{example}{section}

\renewcommand{\oddsidemargin}{0mm}

\usepackage{setspace}
\definecolor{mydarkgreen}{rgb}{0,0.4,0}

\makeatletter
\def\@makefnmark{}

\makeatother
\usepackage{amsfonts,amssymb}
\usepackage{mathrsfs}

\makeatletter
\newcommand{\colim@}[2]{%
  \vtop{\m@th\ialign{##\cr
    \hfil$#1\operator@font colim$\hfil\cr
    \noalign{\nointerlineskip\kern1.5\ex@}#2\cr
    \noalign{\nointerlineskip\kern-\ex@}\cr}}%
}
\newcommand{\colim}{%
  \mathop{\mathpalette\colim@{\rightarrowfill@\scriptscriptstyle}}\nmlimits@
}
\renewcommand{\varinjlim}{%
  \mathop{\mathpalette\varlim@{\rightarrowfill@\scriptscriptstyle}}\nmlimits@
}
\renewcommand{\varprojlim}{%
  \mathop{\mathpalette\varlim@{\leftarrowfill@\scriptscriptstyle}}\nmlimits@
}

\begin{document}
\vspace{-15in}	
	\title{\LARGE {\textbf{Hypothesis testing for medical imaging analysis via the smooth Euler characteristic transform}}}
	\author[1,*]{Jinyu Wang}
    \author[2]{Kun Meng}
    \author[3]{Fenghai Duan}

	\affil[1]{\small Data Science Initiative, Brown University, RI, USA}
    \affil[2]{\small Division of Applied Mathematics, Brown University, RI, USA}
    \affil[3]{\small Department of Biostatistics and Center for Statistical Sciences, RI, USA}
\affil[*]{Corresponding Author: e-mail: \texttt{jinyu\_wang@brown.edu}.}
	
	\maketitle

\begin{abstract}

Shape-valued data are of interest in applied sciences, particularly in medical imaging. In this paper, inspired by a specific medical imaging example, we introduce a hypothesis testing method via the smooth Euler characteristic transform to detect significant differences among collections of shapes. Our proposed method has a solid mathematical foundation and is computationally efficient. Through simulation studies, we illustrate the performance of our proposed method. We apply our method to images of lung cancer tumors from the National Lung Screening Trial database, comparing its performance to a state-of-the-art machine learning model. \footnote{
\begin{itemize}
\item \textbf{Keywords:} Shape-valued data analysis, persistent homology, null hypothesis test, permutation test, medicine.
\end{itemize}
}
\end{abstract}


\section{Introduction}\label{Introduction}

Medical imaging plays a pivotal role in modern healthcare, offering invaluable insights into diagnosing, treating, and monitoring various medical conditions. Among these, images of tumors hold particularly crucial importance. These images, generated through advanced techniques such as computed tomography (CT), magnetic resonance imaging (MRI), and positron emission tomography (PET), provide clinicians with a visual window into the internal structures of the human body, enabling early detection and accurate characterization of tumors.

Statistical inference in medical imaging \citep{bankman2008handbook} holds immense importance as it provides the means to draw meaningful conclusions from image data. It validates findings obtained by medical professionals, ensuring they are not the result of chance.

\begin{figure}[h]
    \centering
    \includegraphics[scale=0.8]{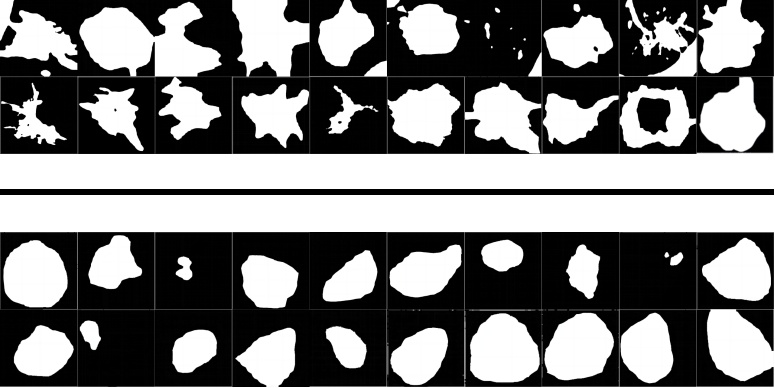}
    \caption{The first two rows represent malignant nodules and the last two rows represent benign nodules.}
    \label{fig: cancer}
\end{figure}

\subsection{Motivating Questions}\label{section: Motivating Questions}

By analyzing CT imaging, medical professionals can often distinguish between malignant and benign tumors. Malignant tumors tend to exhibit certain features on CT images that can help differentiate them from benign tumors. These features may include irregular shapes and poorly defined edges. For example, the (binary segmented) CT scans presented in Figure \ref{fig: cancer} are from the National Lung Screening Trial database \citep{Maldonado2002485}. Each of the scans has been labeled by medical professionals as either malignant or benign. One may ask: ``Do the shapes of malignant tumors differ significantly from those of benign tumors?" A subsequent question is: ``How much statistical power do we have in distinguishing the malignant and benign groups?" These questions motivate our study in this paper.

Many statistical inference methods can be used to address the motivating questions, but efficacy can be compromised by the imposition of stringent model assumptions, especially parametric ones. For example, logistic regression estimates the probability of malignancy using CT features. However, the method's presumption of linearity and independence may not universally apply. Bayesian methods generally depend on and are sensitive to prior distributions.

To address the motivating questions and overcome the referred issues the statistical inference methods suffer from, in this paper, we will propose a distribution-free method via topological data analysis (TDA) \citep{carlsson2009topology}.

\subsection{Overview of Topological Data Analysis and Its Applications}\label{section: Overview of TDA}

TDA is a rapidly growing field that applies algebraic and computational topology to analyze complex data sets  \citep{hatcher2002algebraic, edelsbrunner2022computational, wasserman2018topological}. One of the central techniques in TDA is persistence homology \citep{edelsbrunner2000topological}. The application of persistence homology has proven to be highly effective in a wide range of applications: in bioimage informatics, persistence homology has been used to extract topological information from the single molecule localization microscopy data \citep{Pike400275}; in phylogenetics, persistence homology has been used to analyze the physical features that best describe the variation between mandibular molars from four different suborders of primates \citep{wang2021statistical, meng2022randomness}; persistence homology has also been applied to radio-genomics \citep{crawford2020predicting} and morphology \citep{turner2014persistent, marsh2022detecting}.

Based on persistent homology, \cite{crawford2020predicting} proposed the smooth Euler characteristic transform (SECT) to analyze glioblastoma multiforme. The mathematical foundations of SECT were developed from the Euler calculus viewpoint by \cite{ghrist2018persistent} and from the random fields viewpoint by \cite{meng2022randomness}. Specifically,  \cite{ghrist2018persistent} showed that SECT does not lose any information on shapes when it transforms shape-valued data into functional data;  \cite{meng2022randomness} showed that SECT could be viewed as a random variable taking values in a separable Banach space. 

Shape-valued data are of interest in many applied sciences, e.g., the tumors discussed in the motivating questions in Section \ref{section: Motivating Questions}. The analysis of shape-valued data is still underdeveloped, while functional data analysis has been well developed \citep{wang2016functional}. Using persistent homology, we may transform shape-valued data into functional data, which allows us to implement many tools in functional analysis \citep{brezis2011functional} and functional data analysis.

\subsection{Major Contributions and Paper Organization} \label{section: Major Contributions}

In this paper, we apply the SECT to transform shape-valued data into functional data. Then, we apply the null hypothesis significance testing (NHST, \cite{article}) to the functional data. This approach was motivated by \cite{meng2022randomness}. The combination of SECT and NHST provides us with an effective approach to inferring the differences between two collections of shapes (e.g., the malignant and benign collections of tumors presented in Figure \ref{fig: cancer}).

This paper is organized as follows. In Section \ref{sec:meth}, we provide an overview of the existing background theory for SECT and our proposed hypothesis testing algorithm. We then present proof-of-concept simulation studies in Section \ref{sec:simulations} to show the performance of our proposed algorithm. In Section \ref{sec:app}, we apply our algorithm to the images of lung cancer tumors from the National Lung Screening Trial database. Therein, we compare the performance of our algorithm with a state-of-the-art machine learning approach. A conclusion of this paper is given in Section \ref{sec:conc}.

\section{Theory and Method}\label{sec:meth}

In this section, we first provide a brief description of SECT (Section \ref{subsec:Def}). Then, we propose our hypothesis testing approach (Section \ref{subsec:Rand}).

\subsection{Smooth Euler Characteristic Transform}\label{subsec:Def}

\begin{figure}[h]
    \centering
\includegraphics[scale=0.2]{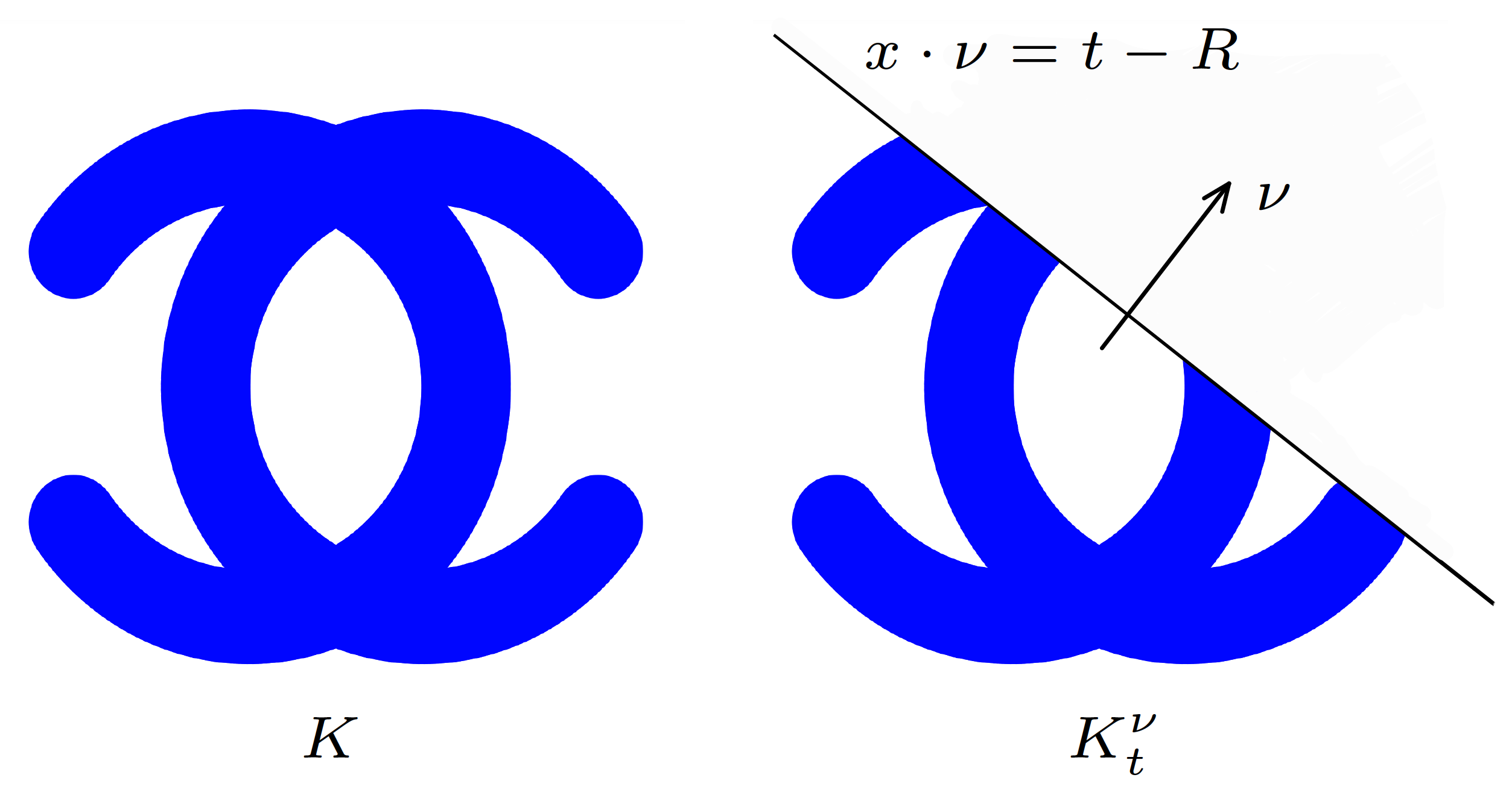}
    \caption{The shape in the left panel is denoted by $K$. For each direction $\nu\in\mathbb{S}^{d-1}$ and scalar $t\in[0,2R]$, the straight line (or hyperplane for high dimensional scenarios) represented by linear equation $x\cdot\nu=t-R$ is associated with them. Then, $K_t^\nu$ is the collection of points in the $K$ that are below this straight line. }
    \label{fig: SECT_illustration}
\end{figure}

Let $K$ denote a compact subset of $d$-dimensional space $\mathbb{R}^d$ (referred to as a ``shape" hereafter), e.g., the tumors in Figure \ref{fig: cancer} and the shape in Figure \ref{fig: SECT_illustration}. We assume the shapes of interest are bounded by a ball with a finite radius, i.e., $K\subseteq B(0,R)$, where $B(0,R)=\{ x\in\mathbb{R}^d : \Vert x\Vert < R\}$ denotes the open ball centered at the origin with some radius $R>0$. For each direction represented by a unit vector $\nu\in\mathbb{S}^{d-1}=\{x\in\mathbb{R}^d \,:\, \Vert x\Vert=1\}$ and scalar $t\in[0,2R]$, we define the following subset of $K$
\begin{align}\label{eq: def of K_tv}
K_t^\nu\overset{\operatorname{def}}{=}\left\{x\in K : x\cdot\nu\le t-R\right\}.
\end{align}
One example illustrating the $K_t^\nu$ defined in Eq.~\eqref{eq: def of K_tv} is presented in Figure \ref{fig: SECT_illustration}. For succinctness, we denote $T\overset{\operatorname{def}}{=}2R$ hereafter.

Let $\chi(K_t^\nu)$ denote the Euler characteristic of $K_t^\nu$. If the subset $K_t^\nu$ is a discrete polygon mesh in computer graphics, we have $\chi(K_t^{\nu})=\#V-\#E+\#F$, where $\#V$, $\#E$, and $\#F$ are the numbers of vertices, edges, and faces of $K_t^\nu$, respectively. Take a triangle (including its face enclosed by its three edges) as an example; the numbers of vertices, edges, and faces of a triangle are 3, 3, and 1, respectively. Therefore, the Euler characteristic of a triangle is 1. When $K_t^\nu$ is not a polygon mesh, the general definition of its Euler characteristic is available in \cite{hatcher2002algebraic}. Efficient algorithms and examples for computing Euler characteristics in statistical applications are available in the literature (e.g., \cite{crawford2020predicting} and \cite{wang2021statistical}). The SECT of shape $K$, denoted as $\operatorname{SECT}(K)$, is a function defined on the Cartesian product space $\mathbb{S}^{d-1}\times [0,T]$. Specifically, $\operatorname{SECT}(K)$ is defined by the following
\begin{align*}
    & \operatorname{SECT}:\ \ K \mapsto \operatorname{SECT}(K)=\left\{\operatorname{SECT}(K)(\nu,t) \right\}_{(\nu,t)\in \mathbb{S}^{d-1}\times [0,T]}, \\
    &\text{where }\ \ \ \operatorname{SECT}(K)(\nu;t) \overset{\operatorname{def}}{=} \int_0^t \chi(K_{\tau}^\nu) \,d\tau-\frac{t}{T}\int_0^T \chi(K_{\tau}^\nu) \,d\tau.
\end{align*}
Visualizations of the $\operatorname{SECT}(K)(\nu;t)$ defined above are presented in Figures \ref{fig: simulation visualizations} and \ref{fig: sect_example}. Under some topological conditions, \cite{ghrist2018persistent} showed that the shape-to-function transform $K \mapsto \operatorname{SECT}(K)$ is invertible, i.e., $\operatorname{SECT}(K)$ contains all the information of shape $K$; in addition, if the shape $K$ is random, \cite{meng2022randomness} showed that $\operatorname{SECT}(K)$ is a random variable taking values in a separable Banach space.

Using the mathematical framework developed in \cite{meng2022randomness}, one can easily show that $\operatorname{SECT}(K)$ belongs to $C(\mathbb{S}^{d-1}; L^2(0,T))=$ the collection of all $L^2(0,T)$-valued continuous functions defined on sphere $\mathbb{S}^{d-1}$, where the function space $C(\mathbb{S}^{d-1}; L^2(0,T))$ is a Banach space equipped with the norm 
\begin{align*}
    \Vert s\Vert_{C(\mathbb{S}^{d-1}; L^2(0,T))} \overset{\operatorname{def}}{=} \sup_{\nu\in\mathbb{S}^{d-1}}\left\{ \left(\int_{0}^T \left\vert s(\nu,t) \right\vert^2 dt\right)^{1/2}\right\}.
\end{align*}
Suppose two shapes $K^{(1)}$ and $K^{(2)}$ are given. As aforementioned, \cite{ghrist2018persistent} proved that all the information about each shape $K^{(i)}$ is contained by the corresponding $\operatorname{SECT}(K^{(i)})$. Hence, the dissimilarity between $K^{(1)}$ and $K^{(2)}$ can be measured by that between the two functions $\operatorname{SECT}(K^{(1)})(\nu,t)$ and $\operatorname{SECT}(K^{(2)})(\nu,t)$. Specifically, we measure the dissimilarity between two shapes by the following
\begin{align}\label{eq: def of rho}
 \begin{aligned}
     \rho\left(K^{(1)}, K^{(2)}\right) & \overset{\operatorname{def}}{=} \left\Vert \, \operatorname{SECT}(K^{(1)}) - \operatorname{SECT}(K^{(2)}) \, \right\Vert_{C(\mathbb{S}^{d-1}; L^2(0,T))}\\ 
     &= \sup_{\nu\in\mathbb{S}^{d-1}}\left\{ \left(\int_{0}^T \left\vert\, \operatorname{SECT}(K^{(1)})(\nu;t) - \operatorname{SECT}(K^{(2)})(\nu;t) \,\right\vert^2 dt \right)^{1/2} \right\}
 \end{aligned}
\end{align}
Our proposed hypothesis testing approach will be based on the dissimilarity function $\rho(\cdot,\cdot)$ defined in Eq.~\eqref{eq: def of rho}.

\subsection{Randomization-style Null Hypothesis Significance Test via the Smooth Euler Characteristic Transform}
\label{subsec:Rand}

The permutation test \citep{good2013permutation} is a non-parametric statistical method that makes no distributional assumptions, making it versatile and applicable to a wide range of data sets. It is especially valuable for small sample sizes (e.g., the size of the malignant/benign tumor collection in Figure \ref{fig: cancer} is only 20) where traditional tests might lack power or their assumptions may not hold. Through rearranging observed data to assess the null hypothesis, the test offers an intuitive and exact approach, controlling the type I error rate effectively. In this subsection, we propose an algorithm for distinguishing two groups of shapes (e.g., the malignant and benign groups of tumors) via the permutation test. 

Suppose there are two underlying shape-generating distributions, $\mathbb{P}^{(1)}$ and $\mathbb{P}^{(2)}$. Shape collections $\{K_i^{(1)}\}_{i=1}^{n_1}$ and $\{K_i^{(2)}\}_{i=1}^{n_2}$ are generated from $\mathbb{P}^{(1)}$ and $\mathbb{P}^{(2)}$, respectively, where the sample size $n_1$ is not necessarily equal to $n_2$ (i.e., unbalanced groups are allowed). We propose a SECT-based randomization-style null hypothesis significance test (NHST-SECT) to test the following hypotheses using collections $\{K_i^{(1)}\}_{i=1}^{n_1}$ and $\{K_i^{(2)}\}_{i=1}^{n_2}$
\begin{align}\label{eq: hypotheses for randomization-style NHST}
H_0:\ \ \mathbb{P}^{(1)}=\mathbb{P}^{(2)}\ \ \ vs.\ \ \ H_1:\ \ \mathbb{P}^{(1)}\ne\mathbb{P}^{(2)}.
\end{align}
Distinguishing the shape collections $\{K_i^{(1)}\}_{i=1}^{n_1}$ and $\{K_i^{(2)}\}_{i=1}^{n_2}$ is represented as rejecting the null hypothesis $H_0$ in Eq.~\eqref{eq: hypotheses for randomization-style NHST}. The randomization-style NHST-SECT is based on permutation test \citep{good2013permutation} and the following loss functional
\begin{align}\label{eq: randomization-style NHST-SECT}
\begin{aligned}
    F\left(\{K_i^{(1)}\}_{i=1}^{n_1}, \, \{K_i^{(2)}\}_{i=1}^{n_2}\right) \overset{\operatorname{def}}{=} \sum_{j=1}^2 \frac{1}{2n_j(n_j-1)} \sum_{k,l=1}^{n_j} \rho\left(K_k^{(j)},\, K_l^{(j)}\right),
\end{aligned}
\end{align}
where $\rho$ is the distance function defined in Eq.~\eqref{eq: def of rho}. 

In applications, it is infeasible to calculate $\operatorname{SECT}(K_i^{(j)})(\nu,t)$ for all infinitely many directions $\nu\in\mathbb{S}^{d-1}$ and levels $t\in[0,T]$. Hence, we calculate the following discretized version in applications
\begin{align}\label{eq: discretized SECT}
    \left\{\operatorname{SECT}(K_i^{(j)})(\nu_p;t_q):p=1,\cdots,\Gamma \mbox{ and } q=1,\cdots,\Delta \right\}_{i=1}^{n_j},\ \ \text{ for }j\in\{1,2\},
\end{align}
where $\{\nu_p\}_{p=1}^\Gamma$ and $\{t_q\}_{q=1}^\Delta$ are prespecified. We choose $\{\nu_p\}_{p=1}^\Gamma$ uniformly from $\mathbb{S}^{d-1}$ and $\{t_q\}_{q=1}^\Delta$ uniformly from $[0,T]$, which is a standard approach in the literature, e.g., \cite{crawford2020predicting}, and will be implemented in Section \ref{subsec:lung}. The discretized SECT values in Eq.~\eqref{eq: discretized SECT} are the input data for our statistical inference.

Given the discretized SECT in Eq.~\eqref{eq: discretized SECT} as our data, we estimate the distance $\rho(K_k^{(j)},\, K_l^{(j)})$ by the following
\begin{align}\label{eq: approximate rho-SECT}
 \rho\left(K_k^{(j)},\, K_l^{(j)}\right) \approx \widehat{\rho}\left(K_k^{(j)},\, K_l^{(j)}\right) \overset{\operatorname{def}}{=}  \sup_{p=1,\ldots,\Gamma}\left(\sum_{q=1}^\Delta \left\vert \operatorname{SECT}(K_k^{(j)})(\nu_p;t_q)-\operatorname{SECT}(K_l^{(j)})(\nu_p;t_q)\right\vert^2\right)^{1/2}.
\end{align}
Correspondingly, the value of the loss functional in Eq.~\eqref{eq: randomization-style NHST-SECT} is approximated by the test statistic $\widehat{F}(\{K_i^{(1)}\}_{i=1}^{n_1},  \{K_i^{(2)}\}_{i=1}^{n_2})$ defined as follows (the test statistic is derived by our input data in Eq.~\eqref{eq: randomization-style NHST-SECT})
\begin{align}\label{eq: approximated loss function}
    F\left(\{K_i^{(1)}\}_{i=1}^{n_1}, \, \{K_i^{(2)}\}_{i=1}^{n_2}\right) \approx \widehat{F}\left(\{K_i^{(1)}\}_{i=1}^{n_1}, \, \{K_i^{(2)}\}_{i=1}^{n_2}\right) \overset{\operatorname{def}}{=} \sum_{j=1}^2 \frac{1}{2n_j(n_j-1)} \sum_{k,l=1}^{n_j} \widehat{\rho}\left(K_k^{(j)},\, K_l^{(j)}\right).
\end{align}
Based on the test statistic defined in Eq.~\eqref{eq: approximated loss function}, we apply Algorithm \ref{algorithm: randomization-style NHST} to implement the NHST-SECT.

\begin{algorithm}
\caption{: Randomization-style NHST-SECT}\label{algorithm: randomization-style NHST}
\begin{algorithmic}[1]
\Statex \textbf{Input:} \noindent (i) Discretized SECT of two collection of shapes $\{\operatorname{SECT}(K_i^{(j)})(\nu_p;t_q):p=1,\cdots,\Gamma \mbox{ and } q=1,\cdots,\Delta\}_{i=1}^{n_j}$ for $j\in\{1,2\}$; (ii) desired confidence level $1-\alpha$ with significance $\alpha\in(0,1)$; (iii) the number of permutations $\Pi$.
\Statex \textbf{Output:} \texttt{Accept} or \texttt{Reject} the null hypothesis $H_0$ in Eq.~\eqref{eq: hypotheses for randomization-style NHST}.
\State Apply Eq.~\eqref{eq: approximate rho-SECT} and Eq.~\eqref{eq: approximated loss function} to the original input SECT data and compute the value of the loss $\mathfrak{S}_0 \overset{\operatorname{def}}{=} \widehat{F}(\{K_i^{(1)}\}_{i=1}^n, \, \{K_i^{(2)}\}_{i=1}^n)$.
\State \textbf{for all $k=1,\cdots,\Pi$}, 
\State Randomly permute the group labels $j\in\{1,2\}$ of the input SECT data.
\State Apply Eq.~\eqref{eq: approximate rho-SECT} and Eq.~\eqref{eq: approximated loss function} to the permuted SECT data and compute the value of the loss $\mathfrak{S}_k \overset{\operatorname{def}}{=} \widehat{F}(\{K_i^{(1)}\}_{i=1}^n, \, \{K_i^{(2)}\}_{i=1}^n)$.
\State \textbf{end for}
\State Compute $k^* \overset{\operatorname{def}}{=}[\alpha\cdot\Pi]\overset{\operatorname{def}}{=}$ the largest integer smaller than $\alpha\cdot\Pi$. 
\State \texttt{Reject} the null hypothesis $H_0$ if $\mathfrak{S}_0<\mathfrak{S}_{k^*}$ and report the output.
\end{algorithmic}
\end{algorithm}

\section{Experiments Using Simulations}\label{sec:simulations}

This section presents simulation studies that demonstrate the performance of our randomization-style NHST-SECT framework presented in Algorithm \ref{algorithm: randomization-style NHST}. We also compare our proposed Algorithm \ref{algorithm: randomization-style NHST} with the ``ECT-based randomization-style null hypothesis significance test" presented in Algorithm \ref{algorithm: previous NHST} (see Appendix \ref{NHST}), which was proposed in \cite{article} and \cite{meng2022randomness}. 

\begin{figure}[h]
    \centering
    \includegraphics[scale=0.1]{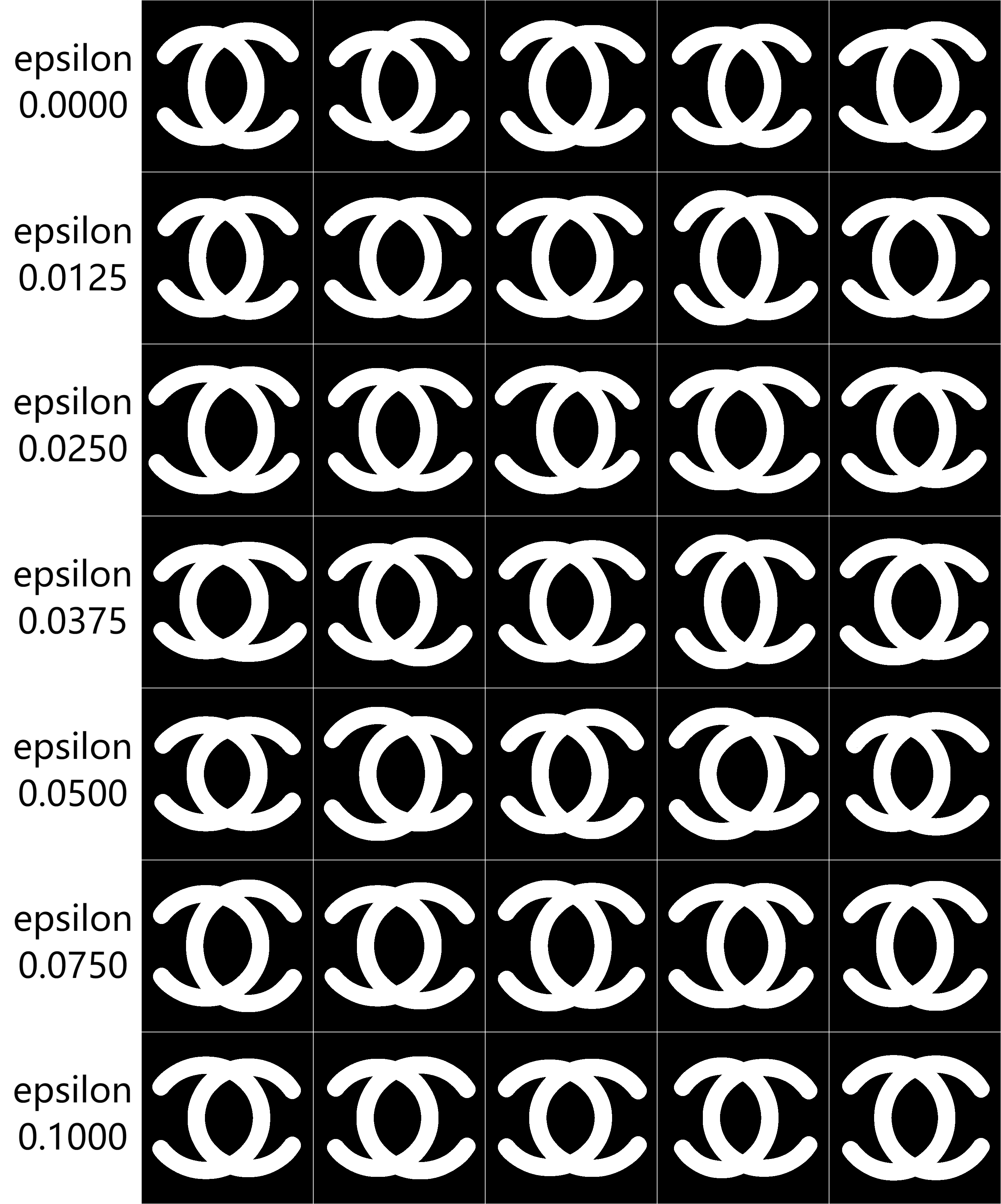}
   \caption{Each row corresponds to an $\varepsilon\in\{0.0000,\, 0.0125,\, 0.0250,\, 0.0375,\, 0.0500,\, 0.0750,\, 0.1000\}$.}
    \label{fig: different_epsilon}
\end{figure}

Our simulation studies focus on a family of distributions $\{\mathbb{P}^{(\varepsilon)}\}_{0\le\varepsilon\le0.1}$, from which we generate a collection of i.i.d. shapes $\{K_i^{(\varepsilon)}\}_{i=1}^n$ for each $\varepsilon\in[0,0.1]$. For each $\varepsilon$, shapes generated from $\mathbb{P}^{(\varepsilon)}$ are presented as follows
\begin{align}\label{eq: explicit P varepsilon}
K_i^{(\varepsilon)}  = & \left\{x\in\mathbb{R}^2 \, \Bigg\vert\, \inf_{y\in S_i^{(\varepsilon)}}\Vert x-y\Vert\le \frac{1}{5}\right\},\ \ \mbox{ where} \\
     \notag   S_i^{(\varepsilon)}  =   & \left\{\left(\frac{2}{5}+a_{1,i}\cdot\cos t, b_{1,i}\cdot\sin t\right) \, \Bigg\vert\, \frac{1-\varepsilon}{5}\pi\le t\le\frac{9+\varepsilon}{5}\pi\right\}\bigcup\left\{\left(-\frac{2}{5}+a_{2,i}\cdot\cos t, b_{2,i}\cdot\sin t\right) \, \Bigg\vert\, \frac{6\pi}{5}\le t\le\frac{14\pi}{5}\right\},
\end{align}
and $\{a_{1,i}, a_{2,i}, b_{1,i}, b_{2,i}\}_{i=1}^n \overset{i.i.d.}{\sim} N(1, 0.05^2)$. The index $\varepsilon$ indicates the dissimilarity between distributions $\mathbb{P}^{(\varepsilon)}$ and $\mathbb{P}^{(0)}$. The varying values of $\varepsilon$ affect the shape generated from $\mathbb{P}^{(\varepsilon)}$. The shapes generated by different $\varepsilon$ across $[0,\,0.1]$ are presented in Figure \ref{fig: different_epsilon}. It is difficult to discern the differences between them by eye. But the TDA-based methods, especially our proposed Algorithm \ref{algorithm: randomization-style NHST}, can distinguish shapes corresponding to different $\varepsilon$ with high confidence. Notably, the images presented in Figure \ref{fig: different_epsilon} resemble the image of a malignant lung cancer tumor (see the circle-like scan in the second row of Figure \ref{fig: cancer}). We then test the following hypotheses using the scheme described in Algorithm \ref{algorithm: randomization-style NHST} 
\begin{align*}
    H_0:\, \mathbb{P}^{(0)}=\mathbb{P}^{(\varepsilon)}\ \ \ vs. \ \ \ H_1:\, \mathbb{P}^{(0)} \ne \mathbb{P}^{(\varepsilon)}.
\end{align*}

\begin{figure}[htb]
    \centering
    \includegraphics[scale=0.35]{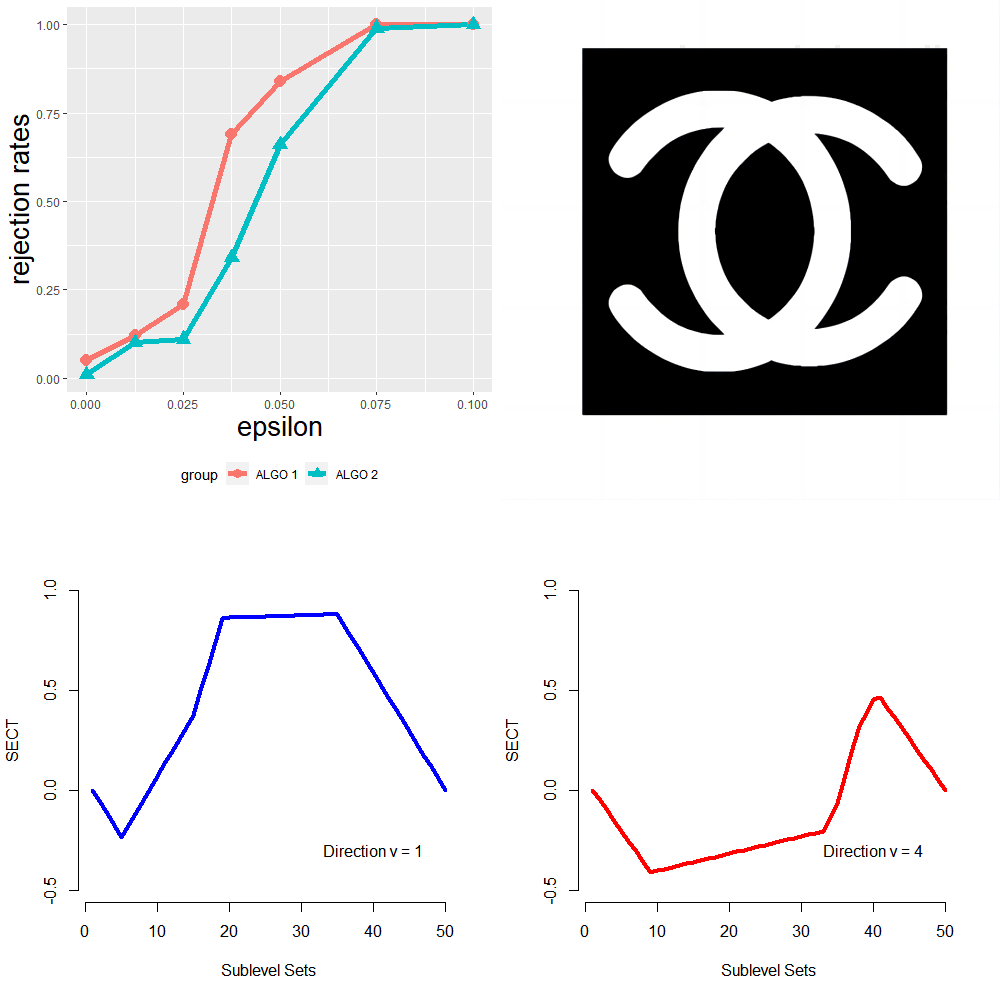}
   \caption{The plot in the top-left panel presents the relationship between $\varepsilon$ and the rejection rates computed via Algorithms \ref{algorithm: randomization-style NHST} (black) and \ref{algorithm: previous NHST} (red) throughout all simulations. The x-axis shows the rejection rates corresponding to $\varepsilon\in\{0.0000,\, 0.0125,\, 0.0250,\, 0.0375,\, 0.0500,\, 0.0750,\, 0.1000\}$. A shape $K$ generated from $\mathbb{P}^{(0.075)}$ is presented in the top-right panel. In the second row presents the curves $t\mapsto \operatorname{SECT}(K)(\nu;t)$ for directions $\nu = (1,0)$ and $(0,1)$, respectively. 
The curves may be interpreted as 1-dimensional manifolds, suggesting a potential correlation between SECT and manifold learning \citep{meng2021principal}.}
    \label{fig: simulation visualizations}
\end{figure}

We conduct experiments by setting $T=3$ and directions $\nu_p=(\cos\frac{p-1}{4}\pi, \sin\frac{p-1}{4}\pi)^T$ for $p\in\{1,2,3,4\}$, sublevel sets with $t_q=\frac{T}{50}q=0.06q$ for $q\in\{1,\cdots,50\}$. We generate two collections of shapes, $\{K_i^{(0)}\}_{i=1}^n\overset{i.i.d.}{\sim} \mathbb{P}^{(0)}$ and $\{K_i^{(\varepsilon)}\}_{i=1}^n\overset{i.i.d.}{\sim} \mathbb{P}^{(\varepsilon)}$, independently for each $\varepsilon\in$ \{0, 0.0125, 0.025, 0.0375, 0.05, 0.075, 0.1\}, using Eq~\eqref{eq: explicit P varepsilon}, with the number of shape pairs set to $n=100$ (here, we take balanced groups as a proof-of-concept example.). We then compute the ECT and SECT of each generated shape in directions $\{\nu_p\}_{p=1}^4$ and at levels $\{t_q\}_{q=1}^{50}$. Next, we implement Algorithms \ref{algorithm: randomization-style NHST} and \ref{algorithm: previous NHST} (with input significance $\alpha=0.05$) on these computed ECT and SECT variables. We repeat this procedure 100 times and report the rejection rates for each epsilon across all 100 replicates in Table \ref{table: epsilon vs. rejection rates}. Moreover, we visually present the rejection rates of Algorithms 1 and \ref{algorithm: previous NHST} in Figure \ref{fig: simulation visualizations}.

\begin{table}[h]
\centering
\caption{Rejection rates of Algorithms \ref{algorithm: randomization-style NHST} and  \ref{algorithm: previous NHST} across different indices $\varepsilon$. We assess the type I error rate of the different algorithms when the null model is true (i.e., $\varepsilon$ = 0). We then assess power of these approaches in the other cases.}
    \label{table: epsilon vs. rejection rates}
    \vspace*{0.5em}
\begin{tabular}{|c|c|c|c|c|}
\hline
Indices $\varepsilon$  & 0.000  & 0.0125  & 0.0250  & 0.0375  \\ [2pt]\hline
Rejection rates of ALGO \ref{algorithm: randomization-style NHST} & 0.05 & 0.12 & 0.21 & 0.69 \\ [2pt]\hline
Rejection rates of ALGO \ref{algorithm: previous NHST} & 0.01 & 0.10 & 0.11 & 0.34    \\ [2pt]\hline
Indices $\varepsilon$  & 0.0500 & 0.0750 & 0.100 & 0.100\\ [2pt]\hline
Rejection rates of ALGO \ref{algorithm: randomization-style NHST}
& 0.84 & 1.00 & 1.00 & 1.00 \\ [2pt]\hline
Rejection rates of ALGO \ref{algorithm: previous NHST} & 0.66 & 0.99 & 1.00 & 1.00 \\ [2pt]\hline
\end{tabular}
\end{table}

The simulation results demonstrate the effectiveness of our proposed Algorithm \ref{algorithm: randomization-style NHST} in detecting the differences between $\mathbb{P}^{(\varepsilon)}$ and $\mathbb{P}^{(0)}$. Specifically, as the value of $\varepsilon$ increases, the rejection rates under the alternative hypothesis of Algorithm \ref{algorithm: randomization-style NHST} also increase. And the power of Algorithm \ref{algorithm: previous NHST} under the alternative hypothesis is weaker compared to our proposed Algorithm \ref{algorithm: randomization-style NHST}. These findings highlight the superior performance of Algorithm \ref{algorithm: randomization-style NHST} in detecting deviations between the two shape-generating distributions. In addition, the scenario in Table \ref{table: epsilon vs. rejection rates} where $\varepsilon=0$ indicates that our proposed Algorithm \ref{algorithm: randomization-style NHST} does not suffer from type I error inflation, which is an inherent feature of a permutation test (recall that our input significance is $\alpha=0.05$).

\section{Applications}\label{sec:app}

In this section, we apply our proposed Algorithm \ref{algorithm: randomization-style NHST} on the low-dose chest computed tomography (CT) images in a lung cancer dataset.

\subsection{Application to Lung Cancer Imaging Dataset}
\label{subsec:lung}
\begin{figure}[h]
    \centering
    \includegraphics[scale=0.25]{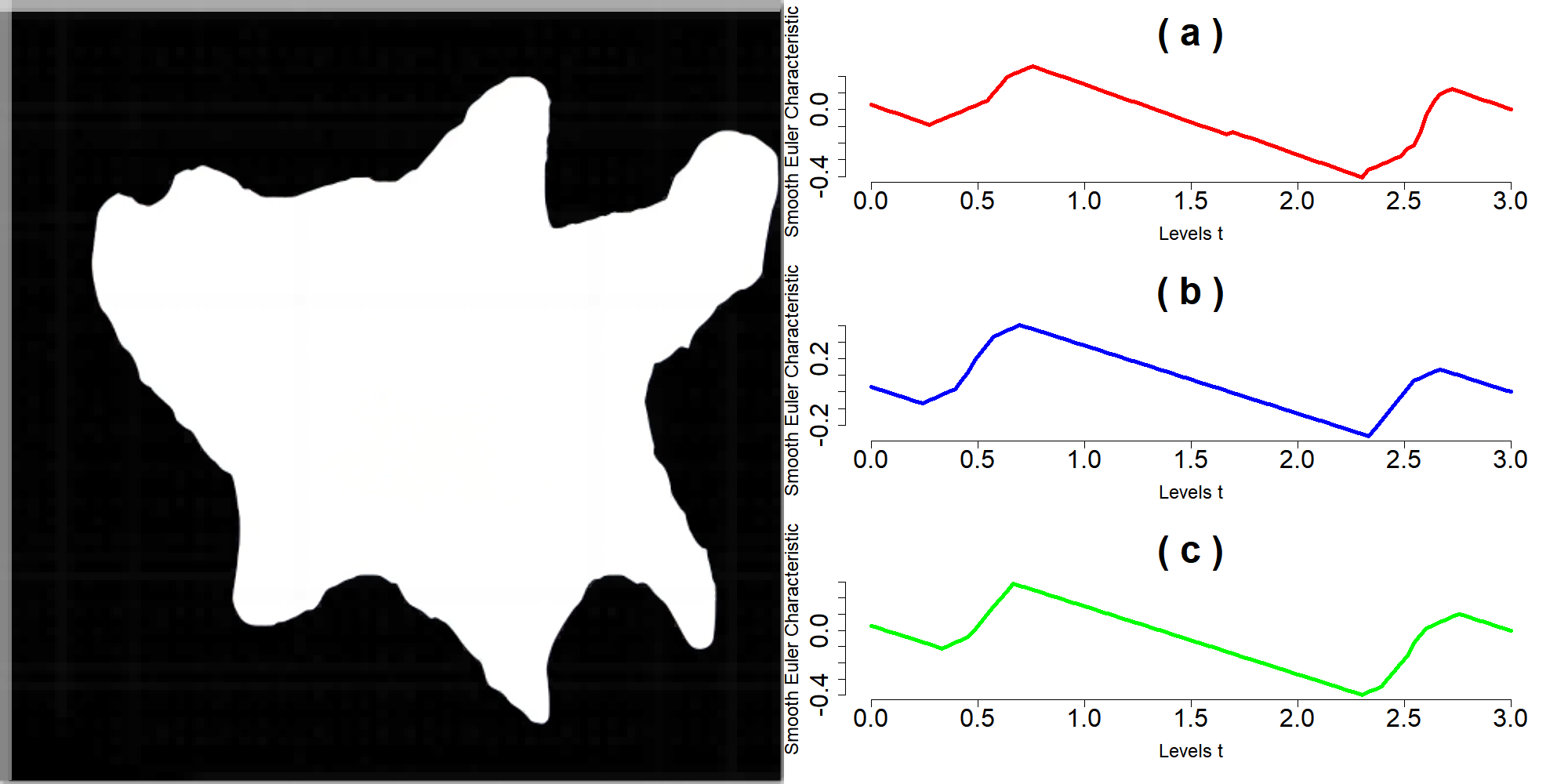}
    \caption{Example of a malignant nodule (left), denoted by $K$, with its SECT curves (right). In the right panel, we illustrate the curves $t\mapsto\operatorname{SECT}(K)(\nu,t)$ for three different directions $\nu$.}
    \label{fig: sect_example}
\end{figure}

CT scans of lung cancer tumors were collected from $n = 685$ patients, archived by National Lung Screening Trial (NLST). Our analysis focused on a subset of this database, distinguishing between two nodule statuses: malignant and benign, as depicted in Figure \ref{fig: cancer}. Each set (malignant or benign) of CT scans comprised 20 grayscale images, representing nodule lesions differentiated from the surrounding lung tissue. Subsequently, these grayscale images underwent thresholding to produce binary images. For each shape $K$ in the data set (i.e., a white region in Figure \ref{fig: cancer}), we compute $\operatorname{SECT}(K)(\nu;t)$ for 72 directions $\nu=(\cos\phi, \sin\phi)$ with $\phi$ evenly sampled over the interval $[0, 2\pi]$, using 100 sublevels $t$ in each direction. The code for implementing our Algorithm \ref{algorithm: randomization-style NHST} in this subsection is built upon the materials in the GitHub repository of authors of \cite{crawford2020predicting}. We modified the code to suit the analysis in this section. Example $\operatorname{SECT}$ summary statistics for a malignant nodule are presented in Figure \ref{fig: sect_example}. To test the hypothesis that malignant tumors exhibit significant differences compared to benign tumors, we utilized Algorithm \ref{algorithm: randomization-style NHST}, with the result presented in Table \ref{tab: cancer} (row 1) by a permutation p-value. Moreover, each dataset was randomly partitioned into two halves, and Algorithm \ref{algorithm: randomization-style NHST} was subsequently applied to detect differences between the two halves. This procedure of random division was conducted 100 times, and the associated p-values, along with their mean and standard deviation, are presented in Table \ref{tab: cancer} (rows 4-5).

The results of implementing Algorithm \ref{algorithm: randomization-style NHST}, as outlined in Table \ref{tab: cancer} (row 1), underscore the capability of our proposed Algorithm \ref{algorithm: randomization-style NHST} to differentiate between distinct nodule statuses. Conversely, rows 4-5 of Table \ref{tab: cancer} suggest that our algorithm refrains from erroneously distinguishing identical nodule statuses.

\begin{table}[h]
\caption{P-values of Algorithm 1 for the lung cancer imaging dataset.}
\label{tab: cancer}
\centering
\begin{tabular}{|c|c|}
\hline

Malignant vs. Benign (Algorithm \ref{algorithm: randomization-style NHST})                       & 0.008    \\[2pt]
Malignant vs. Benign (ResNet18, logistic regression)                        & 0.013    \\[2pt]
Malignant vs. Benign (ResNet18, $\chi^2$-test)                        &  0.024    \\[2pt]\hline
 Malignant vs. Malignant & 0.499 (0.300) \\[2pt]
 Benign vs. Benign&   0.511 (0.306)              \\ [2pt]\hline
\end{tabular}
\end{table}

\subsection{Comparative Analysis of the NHST-SECT Algorithm and Conventional Artificial Neural Network Model}
\label{sec:Comp}

\begin{figure}[h]
    \centering
    \includegraphics[scale=0.6]{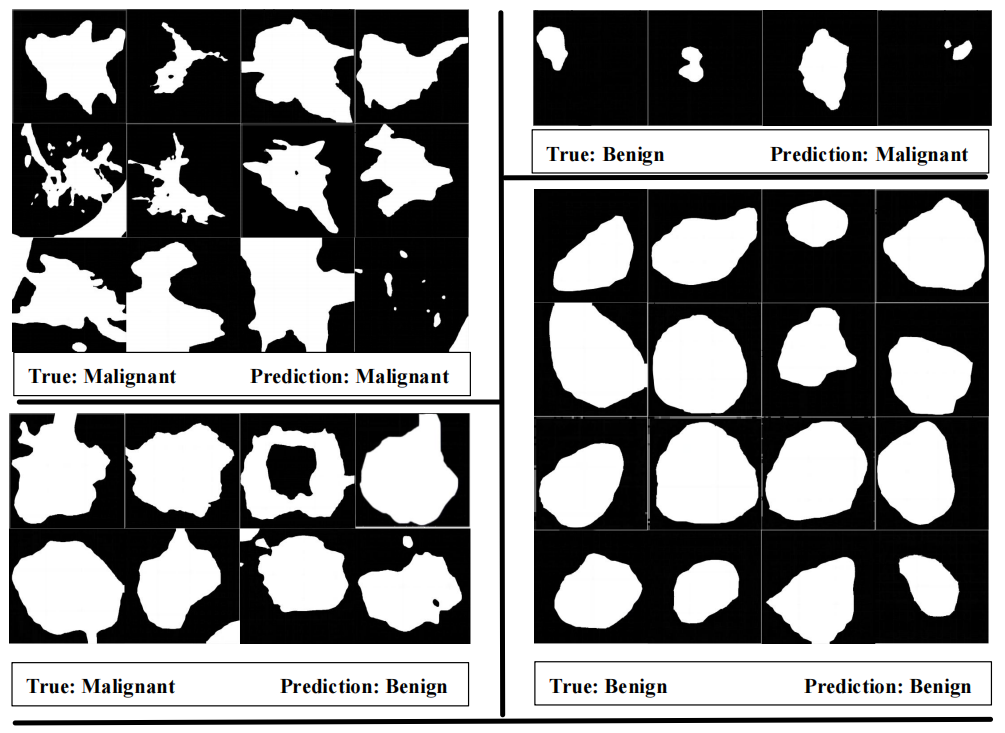}
    \caption{The prediction results of ResNet18 model.}
    \label{fig: prediction}
\end{figure}

In this subsection, we compare our proposed NHST-SECT method (Algorithm \ref{algorithm: randomization-style NHST}) to a state-of-the-art artificial neural network model, focusing on their performance in the lung cancer images presented in Figure \ref{fig: cancer}. We chose the widely used ResNet18 architecture implemented in PyTorch \citep{He2015DeepRL}. We adopted a leave-one-out approach. Specifically, (i) we trained the ResNet18 model on a dataset consisting of 39 CT images, with one image held out for testing in each iteration; (ii) subsequently, the trained model was applied to the held-out image to predict the malignant/benign label of the image (recall that the true label is available); (iii) we iterated this procedure 40 times, with a distinct image held out in each iteration. Consequently, after completing the 40 iterations, each individual image possessed two labels denoting its malignancy/benignity status: the true label and the label predicted through the application of the corresponding trained ResNet18 model. The prediction results via ResNet18 can be found in Figure \ref{fig: prediction}, presented in the form of a $2\times2$ contingency table. We firstly employed logistic regression, regressing the true labels against the predicted labels, subsequently obtaining the p-value associated with the regression coefficient. We also conducted $\chi^2$-test (see \cite{shao2003mathematical}, Section 6.4.3 therein) and obtained the corresponding $\chi^2$-derived p-value. The ResNet18-derived p-values (0.013 for logistic regression, and 0.024 for $\chi^2$-test) are presented in rows 2-3 of Table \ref{tab: cancer}, indicating that our proposed NHST-SECT method is comparable with the state-of-the-art ResNet18 model.

Our proposed Algorithm \ref{algorithm: randomization-style NHST} enjoys several notable advantages over conventional artificial neural network models for medical image classification:
\begin{enumerate}
    \item  Interpretability and Mathematical Foundation. Unlike the black-box nature of the artificial neural network models, our method is grounded in sound mathematical and statistical theories, providing a clear and interpretable basis. 
    \item  Efficient Computational Performance. Efficiency is a critical consideration in medical image analysis, and our method offers reasonable computational efficiency. With a time complexity of $O(n)$, where $n$ represents the number of medical images, our algorithm completed all the data analyses presented in Section \ref{subsec:lung} in approximately 50 seconds (on a computer with an AMD Ryzen 7 5800H processor running at 3200 MHz using 16 GB of RAM). This efficiency rivals popular artificial neural network models, making our method a viable and competitive alternative.
    \item  Robust Performance on Small Datasets. Medical image datasets are often limited in size due to the precious nature of the data. Conventional artificial neural network models are susceptible to overfitting on such small datasets, leading to erroneous classifications on the test set. Due to the permutation technique (recall the discussion at the beginning of Section \ref{subsec:Rand}), our proposed method does not suffer from small sample sizes.
\end{enumerate}

\section{Conclusion and Future Research}
\label{sec:conc}
We introduced the randomization-style NHST-SECT algorithm (Algorithm \ref{algorithm: randomization-style NHST}) for detecting significance differences between shape collections. Simulation studies showed the promising performance of our algorithm. As proof of concept, we applied our proposed algorithm to a dataset of lung cancer images. 

Similar methods can be potentially applied to analyze human brain functional connectivity \citep{lindquist2008statistical, meng2021population}, e.g., detecting significant differences between two collections of functional connectivity structures, which is left for our future research.

\subsection*{DATA AVAILABILITY STATEMENT}

The source code for implementing the simulation studies is publicly available online in the GitHub repository \href{https://github.com/JinyuWang123/NHST-SECT}{https://github.com/JinyuWang123/NHST-SECT.git.} Due to privacy and confidentiality considerations, the data set utilized in Section \ref{sec:app} cannot be made publicly available.

\newpage

\begin{appendix}

\section*{Appendix: ECT-based Randomization-style Null Hypothesis Significance Test}\label{NHST}

The SECT originated from the Euler characteristic transform (ECT, \cite{turner2014persistent}) defined as follows
\begin{align*}
    \begin{aligned}
        & \operatorname{ECT}:\ \ K \mapsto \{\operatorname{ECT}(K)(\nu;t)\}_{(\nu,t)\in \mathbb{S}^{d-1}\times [0,T]}, \\
    & \text{where }\ \ \ \operatorname{ECT}(K)(\nu;t)=\chi(K_t^\nu).
    \end{aligned}
\end{align*}

In Section \ref{sec:simulations}, we compared our proposed Algorithm \ref{algorithm: randomization-style NHST} with the ``ECT-based randomization-style null hypothesis significance test" (see \cite{meng2022randomness} and Section 5.3 of \cite{article}), which is based on permutation test and the following loss functional
\begin{align}\label{eq: randomization-style NHST}
    \begin{aligned}
   H\left(\{K_i^{(1)}\}_{i=1}^{n_1}, \, \{K_i^{(2)}\}_{i=1}^{n_2}\right) \overset{\operatorname{def}}{=} \sum_{j=1}^2\frac{1}{2n_j(n_j-1)}\sum_{k,l=1}^{n_j}\theta\left(K_k^{(j)},\, K_l^{(j)}\right),
    \end{aligned}
\end{align}
where $\theta$ is the dissimilarity function defined by the following 
\begin{align*}
    \theta\left(K^{(1)}, K^{(2)}\right)=\sup_{\nu\in\mathbb{S}^{d-1}}\left\{ \left(\int_{0}^T \left\vert\, \operatorname{ECT}(K^{(1)})(\nu;t) - \operatorname{ECT}(K^{(2)})(\nu;t) \,\right\vert^2 dt \right)^{1/2} \right\}.
\end{align*}
For a given discretized ECT denoted by $\{ECT(K_i^{(j)})(\nu_p;t_q):p=1,\cdots,\Gamma \mbox{ and } q=1,\cdots,\Delta\}_{i=1}^n$, applying the strategy in Eq.~\eqref{eq: approximated loss function}, we may adopt the following test statistics to approximate the loss functional in Eq.~\eqref{eq: randomization-style NHST}
\begin{align}\label{eq: approximate rho}
\begin{aligned}
   & \theta\left(K_k^{(j)},\, K_l^{(j)}\right) \approx \widehat{\theta}\left(K_k^{(j)},\, K_l^{(j)}\right) \overset{\operatorname{def}}{=} \sup_{p=1,\ldots,\Gamma}\left(\sum_{q=1}^\Delta \left\vert\, \operatorname{ECT}(K_k^{(j)})(\nu_p;t_q)-\operatorname{ECT}(K_l^{(j)})(\nu_p;t_q) \,\right\vert^2\right)^{1/2}, \\
   & H\left(\{K_i^{(1)}\}_{i=1}^{n_1}, \, \{K_i^{(2)}\}_{i=1}^{n_2}\right) \approx \widehat{H}\left(\{K_i^{(1)}\}_{i=1}^{n_1}, \, \{K_i^{(2)}\}_{i=1}^{n_2}\right)  \overset{\operatorname{def}}{=} \sum_{j=1}^2\frac{1}{2n_j(n_j-1)}\sum_{k,l=1}^{n_j}\widehat{\theta}\left(K_k^{(j)},\, K_l^{(j)}\right)
\end{aligned}
\end{align}
The ECT-based randomization-style null hypothesis significance test is implemented using Algorithm \ref{algorithm: previous NHST}.

\begin{algorithm}
\caption{: ECT-based randomization-style null hypothesis significance test}\label{algorithm: previous NHST}
\begin{algorithmic}[1]
\Statex \textbf{Input:} \noindent (i) Discretized ECT of two collection of shapes $\{\operatorname{ECT}(K_i^{(j)})(\nu_p;t_q):p=1,\cdots,\Gamma \mbox{ and } q=1,\cdots,\Delta\}_{i=1}^n$ for $j\in\{1,2\}$;
        (ii) desired confidence level $1-\alpha$ with significance $\alpha\in(0,1)$; (iii) the number of permutations $\Pi$.
\Statex \textbf{Output:} \texttt{Accept} or \texttt{Reject} the null hypothesis $H_0$ in Eq.~\eqref{eq: hypotheses for randomization-style NHST}.
\State Apply Eq.~\eqref{eq: approximate rho} to the original input ECT data and compute the value of the loss $\mathfrak{S}_0 \overset{\operatorname{def}}{=} \widehat{H}(\{K_i^{(1)}\}_{i=1}^n, \, \{K_i^{(2)}\}_{i=1}^n)$.
\State for all $k=1,\cdots,\Pi$, 
\State Randomly permute the group labels $j\in\{1,2\}$ of the input ECT data.
\State Apply Eq.~\eqref{eq: approximate rho} to the permuted ECT data and compute the value of the loss $\mathfrak{S}_k \overset{\operatorname{def}}{=} \widehat{H}(\{K_i^{(1)}\}_{i=1}^n, \, \{K_i^{(2)}\}_{i=1}^n)$.
\State end for
\State Compute $k^* \overset{\operatorname{def}}{=}[\alpha\cdot\Pi]\overset{\operatorname{def}}{=}$ the largest integer smaller than $\alpha\cdot\Pi$.
\State \texttt{Reject} the null hypothesis $H_0$ if $\mathfrak{S}_0<\mathfrak{S}_{k^*}$ and report the output.
\end{algorithmic}
\end{algorithm}
\end{appendix}
\newpage

\bibliography{main}

\end{document}